\documentclass[letterpaper]{article} 
\usepackage{aaai25}  
\usepackage{times}  
\usepackage{helvet}  
\usepackage{courier}  
\usepackage[hyphens]{url}  
\usepackage{graphicx} 
\usepackage{natbib}  
\usepackage{caption} 
\usepackage{algorithm}
\usepackage{algorithmic}
\usepackage{booktabs}
\usepackage{multirow}
\usepackage{amsmath}

\usepackage{soul}
\usepackage{xcolor}

\usepackage{newfloat}
\usepackage{listings}
\DeclareCaptionStyle{ruled}{labelfont=normalfont,labelsep=colon,strut=off} 
\floatstyle{ruled}
\newfloat{listing}{tb}{lst}{}
\floatname{listing}{Listing}
\nocopyright 

\title{Catching Dark Signals in Algorithms: Unveiling Audiovisual and Thematic Markers of Unsafe Content Recommended for Children and Teenagers}

\author {
    Haoning Xue\textsuperscript{\rm 1},
    Brian Nishimine\textsuperscript{\rm 2},
    Martin Hilbert\textsuperscript{\rm 2},
    Drew Cingel\textsuperscript{\rm 2},
    Samantha Vigil\textsuperscript{\rm 2},
    Jane Shawcroft\textsuperscript{\rm 2},
    Arti Thakur\textsuperscript{\rm 2},
    Zubair Shafiq\textsuperscript{\rm 2},
    Jingwen Zhang\textsuperscript{\rm 2}
}

\affiliations {
    \textsuperscript{\rm 1}University of Utah\\
    \textsuperscript{\rm 2}University of California, Davis\\
    haoning.xue@utah.edu, \{bnishimine, hilbert, dcingel, slvigil, jestephens, athakur, zshafiq, jwzzhang\}@ucdavis.edu
}

\usepackage{bibentry}

\begin{document}

\maketitle

\begin{abstract}
The prevalence of short form video platforms, combined with the ineffectiveness of age verification mechanisms, raises concerns about the potential harms facing children and teenagers in an algorithm-moderated online environment. We conducted multimodal feature analysis and thematic topic modeling of 4,492 short videos recommended to children and teenagers on Instagram Reels, TikTok, and YouTube Shorts, collected as a part of an algorithm auditing experiment. This feature-level and content-level analysis revealed that unsafe (i.e., problematic, mentally distressing) short videos (a) possess darker visual features and (b) contain explicitly harmful content and implicit harm from anxiety-inducing ordinary content. We introduce a useful framework of online harm (i.e., explicit, implicit, unintended), providing a unique lens for understanding the dynamic, multifaceted online risks facing children and teenagers. The findings highlight the importance of protecting younger audiences in critical developmental stages from both explicit and implicit risks on social media, calling for nuanced content moderation, age verification, and platform regulation.
\end{abstract}

%

\section{Introduction and Related Work}

The contemporary social media environment, exemplified by short video platforms like Instagram Reels, TikTok, and YouTube Shorts, poses many challenges to child safety. Short video platforms have attracted an increasing audience base: since 2020, TikTok has achieved 50 million daily active users in the United States \cite{sherman_tiktok_2020}; Instagram Reels reached 2.35 billion monthly active users in 2023 \cite{ruby_35_2023}. Despite platform policies prohibiting children under 13 years old from creating social media accounts \cite{instagram_about_2024, tiktok_guardians_2024}, existing age verification mechanisms remain largely ineffective \cite{jarvie_online_2024, pasquale_digital_2022}. From 2019 to 2023, Meta received 1.1 million reports of under-13 Instagram accounts \cite{brodkin_meta_2023}. A recent study revealed that children aged 8 to 11 reported frequently accessing these platforms \cite{starks_childrens_2024}. Such ineffective age verifications currently implemented by the platforms may expose children to unsafe and harmful content \cite{jamaludin_age_2024}.

Children on short video platforms are exposed to significant physical and mental health risks. A report indicates that over 33 million social media posts on Instagram and TikTok contain problematic content for children, such as suicide and eating disorders \cite{corbett_study_2024}. Exposure to short videos on TikTok is linked to a higher level of anxiety and depression among children \cite{carville_tiktoks_2023}. Further, children are likely to become the target of sexual harassment on social media platforms \cite{blunt_children_2024}. Social media platforms like Meta not only serve as an information hub but also deliberately target younger audiences with personalized advertisements \cite{brooks_meta_2023}, raising public concerns about their responsibilities in child protection. For example, a recent lawsuit in New York against Instagram, TikTok, and YouTube for promoting addictive and unsafe content to children exemplifies the legal responsibilities and consequences facing short video platforms \cite{the_guardian_new_2024}.

In addition to existing parental control and self-regulation strategies for child protection \cite{wisniewski_parental_2017}, more robust preventive interventions are needed to create a safer online information environment for children and teenagers \cite{jamaludin_age_2024}. Existing research has demonstrated ways for documenting and quantifying harmful content recommendation and detection. For example, a recent study reveals that children searching for popular videos on YouTube are frequently exposed to harmful content (e.g., disturbing, violent thumbnails) in the search result list \cite{radesky_algorithmic_2024}. In terms of content detection, Kaushal and colleagues \cite{kaushal_kidstube_2016} developed a detector for unsafe content (i.e., with keywords of graphic nudity and abuse) on YouTube relying on video captions and contextual information such as comments and engagement metrics. A more recent work by Tahir and colleagues \cite{tahir_bringing_2020} employed deep learning to detect unsafe content (i.e., fake, explicit, and violent videos) on YouTube based on visual and audio features.

\subsection{Research Gaps and Aims}
Although this line of research showcases the potential of harmful content detection on social media, there remain several important gaps in a quickly evolving field. Here we contribute by addressing two critical research gaps.

First, existing work has narrowly focused on overtly harmful content, such as short videos depicting violence or nudity. However, a large portion of short video content that can potentially trigger mental health issues and harm well-being––intentionally or unintentionally–has not been well studied. Although prior research has touched upon the potential harm to mental health \cite{amnesty_international_driven_2023, scheinbaum_darker_2024}, a systematic analysis of this content is lacking. Therefore, this study addresses this gap by examining unsafe content with two dimensions: \textit{problematic} and \textit{mentally distressing} content. Problematic short videos include content inappropriate for children, such as explicit or violent content, while mentally distressing short videos include content that may induce anxiety or depression among the audience, such as content that implies “there is something wrong with you” \cite{hilbert__2023}. By incorporating both types of unsafe content, we aim to provide a more comprehensive overview of the online information risks that children face.

\begin{figure}[t]
\centering
\includegraphics[width=0.95\columnwidth]{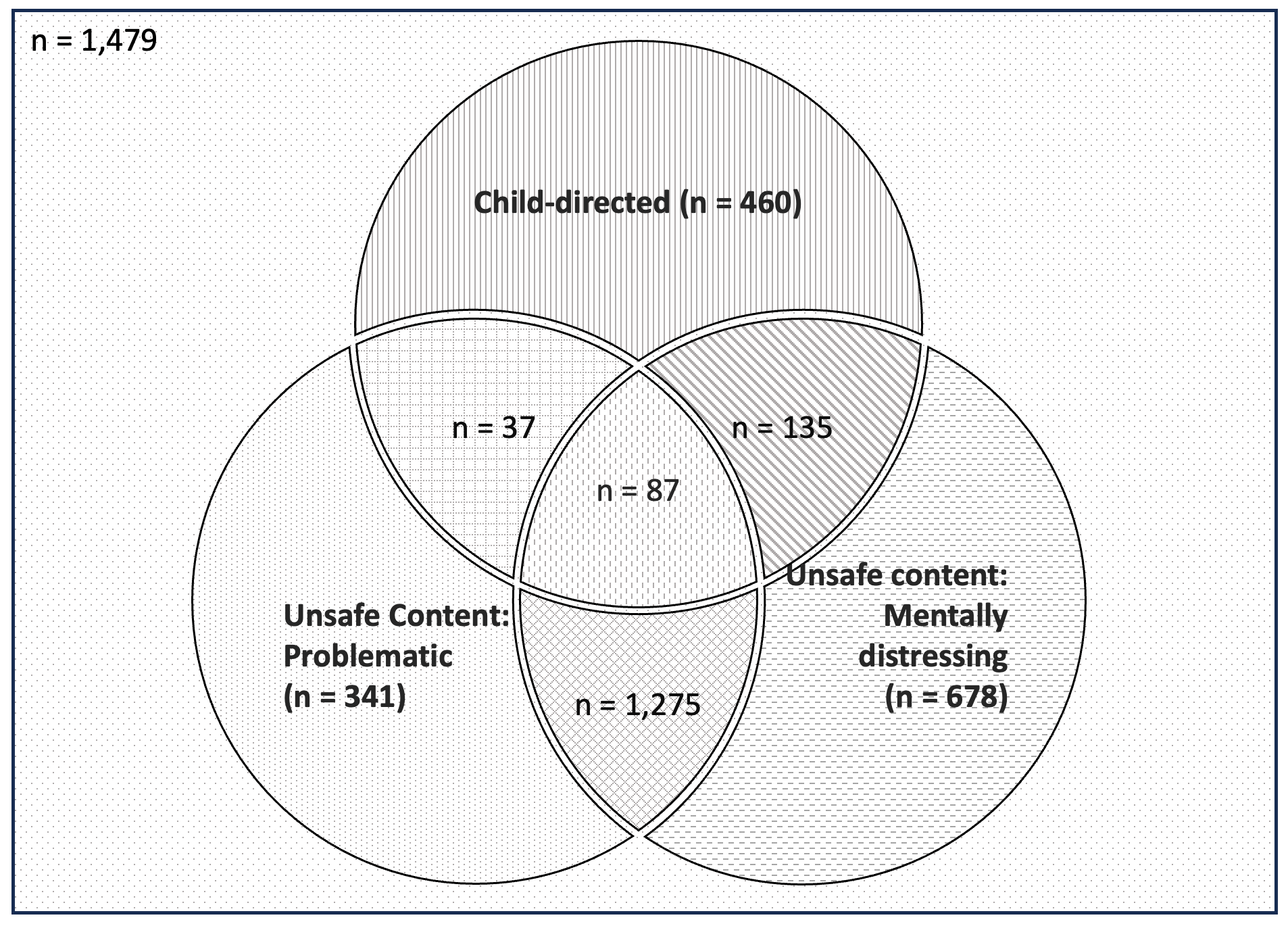}
\caption{Venn diagram of conceptualization of the coding categories of child-directed, problematic, or mentally distressing, and the distribution of short videos across the three coding categories.}
\label{figure1}
\end{figure}

Second, there is a limited understanding of the specific characteristics of video content recommended to children, despite their popularity on the platforms. To fill this gap and provide a comprehensive overview, we dissect the content types into (a) unsafe content recommended to children (i.e., \textit{problematic}, \textit{mentally distressing}) and (b) \textit{child-directed content} (i.e., clearly intended for children) which can be unsafe or safe (see Figure 1 for a Venn diagram illustrating the conceptualization of the content types). Existing research showcases the potential of content detection and moderation with deep learning. Meanwhile, this line of work lacks explainability, offering little insight into how multimodal and thematic features contribute to identifying unsafe and child-directed content. This gap underscores the need for more detailed feature-level and content-level analysis to capture distinctive audiovisual features of unsafe short videos, which could inform the development of more interpretable and effective content detection and moderation strategies.

Relying on a dataset of children’s video recommendations on Instagram Reels, TikTok, and YouTube Shorts, this study explores the audiovisual and thematic markers of child-directed and unsafe content recommended to children and teenagers. Specifically, we aim to answer three research questions. The first question is about the multimodal features of manually annotated (a) child-directed content and (b) unsafe content (see Methods for details of multimodal features). The second research question aims to explore the level of engagement of different types of content, examining how recommendation algorithms prioritize viral content for children and teenagers across different content categories. The last research question aims to identify thematic topics associated with unsafe content, providing a comprehensive overview of recurring themes in unsafe content that can better inform unsafe content detection and moderation. 

\begin{itemize}
\item {\textbf{RQ1a.}} How does child-directed content differ from non-child-directed content on short video platforms regarding multimodal features?
\item {\textbf{RQ1b.}} How does unsafe content (i.e., problematic, mentally distressing) recommended to children and teenagers differ from safe content regarding multimodal features?
\item {\textbf{RQ2.}} How are unsafe content (i.e., problematic, mentally distressing) and child-directed content recommended to children and teenagers associated with video engagement (i.e., the number of views, likes, and comments)?
\item {\textbf{RQ3.}} What thematic topics characterize unsafe content recommended to children and teenagers?
\end{itemize}

To address these questions, we employ a combination of computational content analysis (i.e., multimodal feature analysis, topic modeling) and manual annotation to analyze both the video content and relevant contextual information (i.e., video captions, engagement metrics). We highlight two main findings in this study. First, the negative influence of unsafe short videos manifests through both harmful video content and multimodal features, such as darker visuals. Second, we identified three distinctive ways unsafe content can produce harm: (a) \textit{explicit harm} through inappropriate content, (b) \textit{implicit harm} framed as entertainment and commercialization, and (c) \textit{unintended harm} embedded in mental health-related content.

The contribution of this study can be summarized in two perspectives.

\begin{itemize}
\item First, this study introduces a novel framework for understanding online harms, categorizing unsafe content into three distinct forms: \textit{explicit}, \textit{implicit}, and \textit{unintended} harms. While prior research has primarily focused on overtly explicit harms (e.g., violence, nudity), our study broadens the scope of the research on online harms. This framework provides a unique lens for understanding the dynamic, multifaceted online risks. This detailed approach fills a significant research gap by highlighting how seemingly safe content may cause emotional and psychological distress among children and teenagers. Further, it enhances the explainability of content detection systems by uncovering thematic and multimodal markers associated with unsafe content.
\item Second, our findings provide actionable implications for child protection on social media from both technical and regulatory perspectives. Our findings on the three forms of harm call for more nuanced detection, moderation, and removal strategies. From a policy standpoint, the study’s findings underscore the need for stricter age verification and content regulation to better protect children from the broad spectrum of online harms. By addressing both explicit and implicit risks, this study provides comprehensive empirical evidence for improving platform governance and algorithmic accountability in online content moderation.
\end{itemize}

\section{Methods}
\subsection{Data Collection}

To answer the proposed research questions, we analyzed a publicly available dataset derived from an algorithm-auditing experiment involving short videos recommended to children and teenagers \cite{hilbert__2023} (License: CC BY 4.0). The study created various social media profiles following a factorial design with 3 user groups (age and mental health: 8-year-old, 16-year-old, struggling 16-year-old) × 2 gender (female, male) × 2 race (Black, White), resulting in 12 conditions. In total, 396 unique social media profiles within 12 conditions were created on YouTube, Instagram, and TikTok. Social media content (i.e., images and videos) recommended to the experimental profiles was collected at three time points between April 2023 and May 2023. Among the recommended content, n = 7,920 unique URLs were extracted, including standard YouTube videos, Instagram posts, and short videos from all three platforms. Among these URLs, N = 4,492 URLs identified as short videos were included for the current analysis; standard YouTube videos and Instagram posts were excluded.

We used Apify to download these short videos and retrieve metadata in October 2023 for further analysis \cite{apify_apify_2023}. See Table 1 for summary statistics of the dataset across three platforms. The discrepancy in the number of URLs and the number of videos available can be attributed to certain videos becoming unavailable or set to private during the data collection period.

\begin{table}[ht]
    \centering
    \begin{tabular}{p{0.28\columnwidth}p{0.17\columnwidth}p{0.17\columnwidth}p{0.17\columnwidth}}
        \toprule
        & \textbf{Tiktok} & \textbf{Instagram Reels} & \textbf{YouTube shorts} \\
        \midrule
        \#URLs retrieved & 3,960 & 1,980 & 1,980 \\
        \#URLs of short videos & 2,690 & 810 & 996 \\
        \#Short videos retrieved & 2,658 & 721 & 931 \\
        \midrule
        \#View & 13,732,715 & 10,243,781 & 98,729,873 \\
        \#Likes & 1,143,366 & 1,126,105 & 3,432,550 \\
        \#Comments & 8,110 & 3,786 & 14,351 \\
        \bottomrule
    \end{tabular}
    \caption{Summary statistics of short videos across three short video platforms: Instagram Reels, TikTok, and YouTube Shorts.}
    \label{table1}
\end{table}

\subsection{Multimodal Feature Extraction}

We extracted 16 multimodal features, covering basic audiovisual features that are well-established in aesthetic literature \cite{bannister_vigilance_2020, wyszecki_color_2000}. Further, since human faces are often portrayed in short videos, we included basic face features in this study as well. Although we are aware of the rapid advancement of multimodal features, particularly platform-specific functions and filters, we focus on 16 features that are consistently available across most short videos and platforms to ensure generalizability. The feature extraction methods and descriptive statistics are detailed in Table 2.

\subsubsection{Audio feature extraction} We used the Librosa Library in Python \cite{mcfee_librosa_2015} to extract three audio features: loudness, tempo, and sound brightness. Loudness represents the energy of a sound wave, ranging from 0 to 1. Tempo indicates audio pace, ranging from 0 to positive infinity. Lastly, sound brightness (measured via spectral centroid) represents how bright a sound is perceived, with a higher value indicating a brighter and sharper sound (with more high-frequency content), while a lower value indicates a darker, more bass-heavy sound.

\subsubsection{Face feature extraction} We used Face++ \cite{face_face_2023} to extract the estimated age and six discrete facial expressions (i.e., disgust, anger, fear, happiness, sadness, and surprise) of faces in a video. Face++ is a face recognition tool that can detect faces in an image and analyze demographics and facial expressions. We sampled representative frames in a video using equal interval frame sampling \cite{wang_temporal_2016} and computed the average age and facial expressions for all sampled frames. Facial expressions range from 0 to 100, with higher scores indicating more pronounced facial expressions.

\subsubsection{Visual Feature Extraction} Visual features include motion-level features (i.e., shot changes and face count) and frame-level features (i.e., brightness, color warmness, color saturation, and visual complexity). To extract motion-level features, we used Google Cloud Video Intelligence API \cite{google_cloud_video_2024} to extract the number of shot changes and the number of faces appearing in a video. To extract frame-level features, we used the OpenCV Library in Python \cite{bradski_opencv_2000} to compute the average values of these visual features in sampled frames as an estimation of the overall visual features in a video. First, brightness and saturation are the average values of every pixel in a given frame, ranging from 0 to 255. Higher brightness represents more luminance, while higher saturation represents more vibrance. Second, color warmness is calculated as the proportion of pixels with warm colors in a frame, ranging from 0 to 1. Warm color is indicated by a hue value between 0 (i.e., red) and 30 (i.e., yellow) in the HSV (i.e., hue, saturation, and value) color scheme. Lastly, we used the entropy of pixel textures in a frame to estimate visual complexity \cite{lu_pervasive_2022, qi_improving_2021, yang_using_2019}. It ranges from 0 to 8, with higher values representing more complexity of visual patterns.

\begin{table*}[ht]
  \centering
  \begin{tabular}{cllccccc}
    \toprule
    \# & \textbf{Multimodal feature} & \textbf{Definition} & \textbf{Range} & \textbf{Min} & \textbf{Max} & \textbf{Mean (SD)} \\
    \midrule
    1 & Brightness & Intensity of color luminance & 0, 255 & 0.00 & 258.40 & 122.58 (24.73) \\
    2 & Color saturation & Intensity of color vividness & 0, 255 & 0.00 & 0.52 & 0.06 (0.08) \\
    3 & Color warmness & Percentage of warm colors in all colors & 0, 1 & 0.00 & 4772.08 & 2026.38 (558.05) \\
    4 & Visual complexity & Diversity of visual brightness/texture & 0, 8 & 0.18 & 254.02 & 112.40 (36.42) \\
    5 & Shot count/s & Number of shot changes per second & 0, +$\infty$ & 0.00 & 254.74 & 73.88 (34.12) \\
    6 & Shot duration/s & Average duration per shot per second & 0, 1 & 0.04 & 7.80 & 6.77 (1.15) \\
    7 & Loudness & Energy of a sound wave & 0, 1 & 0.00 & 0.97 & 0.28 (0.19) \\
    8 & Tempo & Sound rhythm, beats per minute & 0, +$\infty$ & 0.00 & 482.00 & 18.95 (27.48) \\
    9 & Sound brightness & Brightness of a sound wave & 0, +$\infty$ & 0.37 & 502.95 & 12.48 (29.47) \\
    10 & Face count/s & Number of faces present per second & 0, +$\infty$ & 0.00 & 1945.00 & 30.55 (69.75) \\
    11 & Face duration/s & Average duration per face per second & 0, 1 & 0.30 & 248.87 & 4.90 (11.16) \\
    12 & Face age & Detected age of faces in a video & 0, +$\infty$ & 2.20 & 82.50 & 31.24 (9.99) \\
    13 & Disgust & Disgust & 0, 100 & 0.00 & 99.99 & 6.09 (11.79) \\
    14 & Anger & Anger & 0, 100 & 0.00 & 99.70 & 5.33 (10.31) \\
    15 & Fear & Fear & 0, 100 & 0.00 & 99.92 & 4.06 (8.75) \\
    16 & Happiness & Happiness & 0, 100 & 0.00 & 100.00 & 20.85 (22.92) \\
    17 & Sadness & Sadness & 0, 100 & 0.00 & 99.99 & 11.33 (15.89) \\
    18 & Surprise & Surprise & 0, 100 & 0.00 & 100.00 & 12.15 (17.37) \\
    \bottomrule
  \end{tabular}
  \caption{Summary statistics of multimodal features in short videos recommended to children and teenagers.}
  \label{table2}
\end{table*}

\subsection{Short Video Annotation}
We employed the annotated data from the publicly available dataset \cite{hilbert__2023}, where 32 human coders assessed each short video on three categories: (a) whether it was directed to children (i.e., clearly intended for children), (b) whether it was problematic for children (i.e., content not suited for children under age 13, such as nudity and violence, according to the Motion Picture Association film rating system), and (c) whether it was mentally distressing (e.g., content implying “there’s something wrong with you” or triggering upward or downward social comparison; see study preregistration for details: \url{https://osf.io/kdbn5/}). Each category was coded as 1 (feature present) or 0 (feature absent or unclear). The final score for each category was the average ratings of the coders. A coding category is considered as present in a video if at least one coder identified it, resulting in a final score larger than 0. The distribution of short videos across these categories is illustrated in Figure 1. Full annotation details are available \cite{hilbert__2023}. On average, short videos in this dataset were rarely child-directed (M = 0.04, SD = 0.11) but were somewhat problematic (M = 0.16, SD = 0.25) and distressing (M = 0.17, SD = 0.23).

\subsection{Analysis Plan}
To answer RQ1 about how multimodal features are presented in (a) child-directed and (b) unsafe content (i.e., problematic, mentally distressing) short videos, we conducted three linear regression models with continuous annotation results as dependent variables , implemented with the R Stats Package. We included 16 multimodal features as independent variables and video length as a covariate. Multimodal features and video length were normalized to ensure comparability.

In addition, we conducted machine learning-based predictive models with XGBoost regression \cite{chen2016xgboost} to evaluate how well multimodal features can predict unsafe content. XGBoost was selected for its robustness with sparse data and non-normal distributions \cite{chen2016xgboost}.

To answer RQ2 about the relationship between content types and video engagement, we conducted three negative binomial regression models using the R MASS Package to account for overdispersion in count data, with continuous annotations of child-directed and unsafe content as independent variables, and video engagement (i.e., the sum of views, likes, and comments) as the dependent variable. We chose linear and binomial regression models to provide interpretable results, allowing us to identify key multimodal features and understand the direction of their influence. The regression coefficients are directly provided by these modeling functions. To complement these estimates in both models, we derived confidence intervals and 95\% confidence intervals using standard profile likelihood methods via the confint() function using the R Stats Package.

\subsubsection{Topic Modeling} To answer RQ3 about the thematic topics for unsafe content, we conducted Latent Dirichlet allocation (LDA) topic modeling \cite{blei_latent_2003} using the Gensim library in Python \cite{rehurek_software_2010}. This analysis was performed on the captions of short videos that have been annotated as problematic (n = 651) or mentally distressing (n = 833). In addition to unsafe content, we performed topic modeling for all available short video captions (n = 1,676) in the dataset to provide a broader thematic overview of all short video content.

Prior to topic modeling, we preprocessed short video captions by removing URLs, emojis, stopwords, and non-English content. The list of stopwords is a combination of the default stopwords from the nltk library in Python \cite{bird_natural_2009} and the top 10 most frequently used words specific to our dataset, including fyp, follow, shorts, video, foryou, subscribe, tiktok, viral, videos, and like. These platform-specific keywords and hashtags were excluded because they are primarily used to increase the visibility of short videos on users’ For You page \cite{burke_tiktok_2024}. However, these keywords do not contribute meaningfully to thematic analysis. After text preprocessing, there were n = 1,676 captions remaining for topic modeling.

\begin{figure*}[t]
\centering
\includegraphics[width=0.95\textwidth]{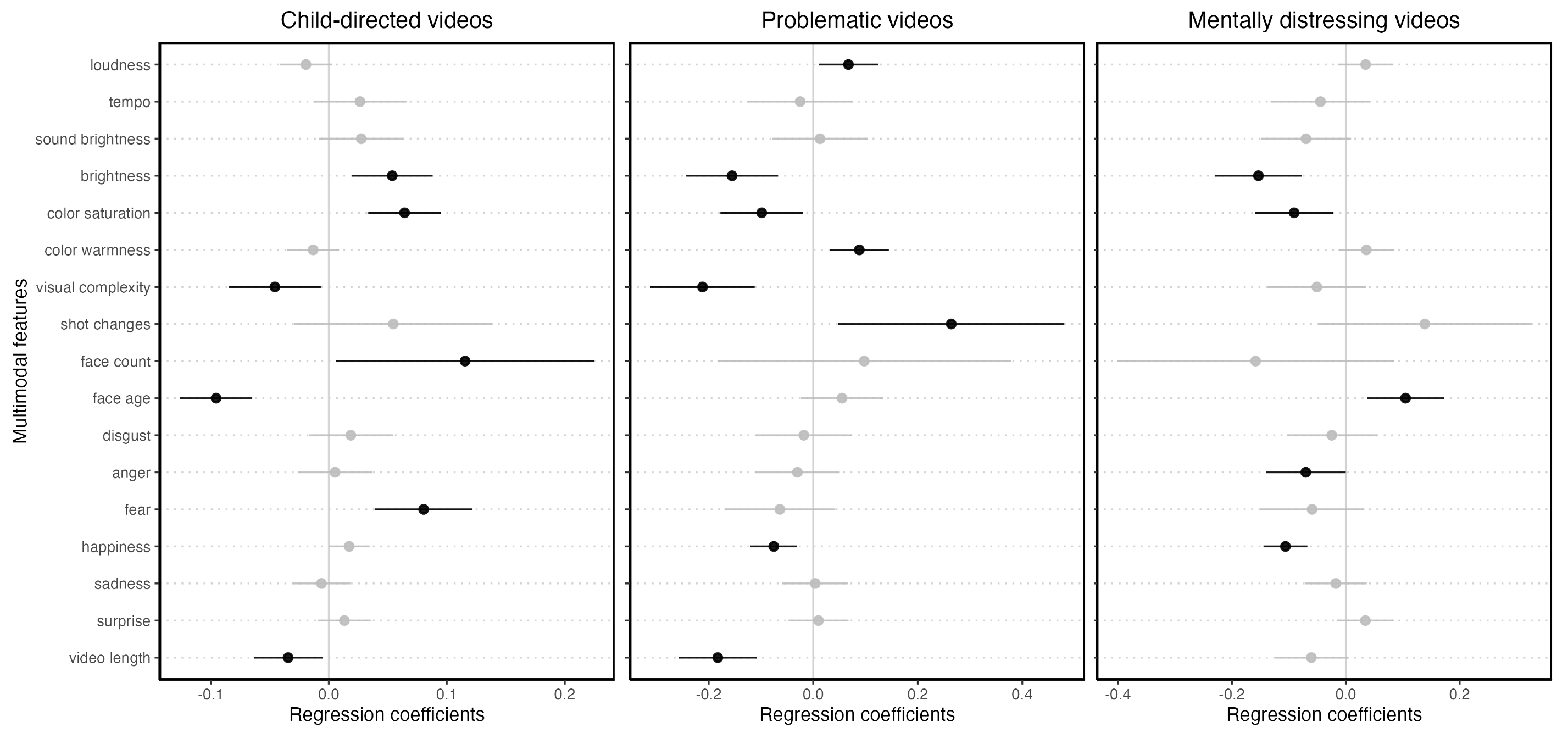}
\caption{Regression coefficients and confidence intervals of multimodal features on the extent to which short videos are (1) child-directed, (2) problematic, or (3) mentally distressing. Multimodal features are normalized for comparison. Regression coefficients with statistically significant effects are marked in black.}
\label{figure2}
\end{figure*}

The topic modeling consists of three key steps. First, we determined the optimal number of topics by running the model with a range of topics from 1 to 10. To select the optimal number, we evaluated the results by balancing two metrics: perplexity and topic coherence, which are commonly used to assess the complexity and interpretability of topic models \cite{rehurek_software_2010}. Second, after identifying the optimal number of topics, we performed the final topic modeling to extract the top 10 keywords contributing to each topic as well as the distribution of these topics for the given documentation. Lastly, since LDA topic modeling relies on word frequency and distribution, it comes with the limitation that repetitive words in a single caption may be overrepresented and therefore potentially distort the results. Given this limitation, the researchers qualitatively evaluated the captions associated with each topic to provide a high-level, accurate summary of thematic topics that emerged from the dataset.

\section{Results}
\subsection{Unsafe Short Videos are Darker and Less Engaging}

RQ1 explored how multimodal features are presented in (a) child-directed content in comparison with non-child-directed content and (b) unsafe content compared with safe content. See Figure 2 for visualization of results. Child-directed short videos tend to be visually brighter, more saturated, and visually simpler, than non-child-directed content. These videos feature a greater number of faces––particularly younger faces with more fearful facial expressions. These bright visuals and young faces can capture children’s attention and effectively engage them. However, multimodal features of problematic and mentally distressing short videos are different. Problematic and distressing short videos use darker visuals with low brightness and saturation, featuring fewer happy facial expressions. Problematic and mentally distressing short videos differ slightly from each other. While problematic videos have warmer tones and maintain visual simplicity, mentally distressing content often features older faces. Across child-directed and unsafe short videos, brightness and saturation consistently emerge as key distinguishing features: child-directed content tends to be brighter and more saturated, while problematic and distressing content is darker and less vibrant.

The distinct multimodal features of unsafe content are also evident in child-directed short videos. Since the annotations of child-directed and unsafe content were conducted independently, we looked deeper into child-directed short videos. Within child-directed short videos (n = 719), unsafe content accounted for 36.0\% (combined n = 259; Figure 1) and exhibited unique multimodal features. Specifically, problematic content directed to children tended to have higher levels of tempo and less warm colors; mentally distressing content directed to children showed higher levels of tempo and lower levels of brightness. Therefore, unsafe content directed to children possessed distinct multimodal features from safe content directed to children and teenagers.

Further, we conducted separate regression models for each short video platform to explore platform-specific trends (see Appendix 1 for visualization). Overall, facial features are more predictive for short videos on Instagram Reels, while audiovisual features are more prominent for TikTok and YouTube Shorts. For instance, child-directed videos on YouTube Shorts possessed brighter visual features, while facial features were more predictive on Instagram Reels. As for problematic and mentally distressing videos, most multimodal features had weaker or even non-significant associations on Instagram Reels except for happy facial expressions, but unsafe videos on TikTok and YouTube Shorts were visually darker. These differences highlight the heterogeneous nature of platform ecosystems, where similar content categories may manifest through divergent audiovisual signals.

Lastly, we conducted predictive models with XGBoost regression to evaluate the predictiveness of multimodal features in distinguishing unsafe content recommended to children. We used grid search to optimize the parameters of the XGBoost models. The models demonstrated strong predictive performance, such that multimodal features can predict problematic content (MSE = $3.504 \times 10^{-5}$, AIC = $-4578.506$, $MSE_{\text{baseline}} = 0.069$) and mentally distressing content (MSE = $2.260 \times 10^{-5}$, AIC = $-4775.857$, $MSE_{\text{baseline}} = 0.055$). These results suggest that multimodal features can effectively predict unsafe content, offering a scalable and cost-efficient approach to improving content moderation systems.

RQ2 asked about the relationship between content types and video engagement. Consistently, child-directed content is more engaging, positively predicting the number of views, likes, and comments. Unsafe content is generally less engaging: although problematic content was not significantly related to engagement metrics, mentally distressing content negatively predicted the number of views and likes. This result suggested that child-directed short videos are generally more engaging than unsafe content with brighter visuals and higher appeals.

\begin{table*}[ht]
  \centering
  \begin{tabular}{p{0.03\textwidth}p{0.05\textwidth}p{0.10\textwidth}p{0.15\textwidth}p{0.56\textwidth}}
    \toprule
    \textbf{Topic \#} & \textbf{Count (\%)} & \textbf{Topic \newline Summary} & \textbf{Top Keywords} & \textbf{Example Captions} \\
    \midrule
    1 & 351 (53.9\%) & Promotion, self-advertisement & \raggedright dog, train, instagram, link, twitter, channel, new, watch, share, daily & 
    1. "on instagram it was last night it’s my most recent post ... is my instagram live daily their follow me..." \newline
    2. "I'm a 27 year old magician... If you enjoyed the magic video and wanna see more dope magic tricks + incredible performances please SUBSCRIBE... \#Shorts \#Magic" \newline
    3. "...If you would like to create a side income while helping people with their nutritional needs DM me the word ‘RESET’ I will reach out" \\
    \midrule
    2 & 136 (20.9\%) & Crime, violence, emotional trauma & \raggedright mentalhealth, join, explorepage, explore, entertainment, katana, carter, edit, inspire, trend & 
    1. "hanuman 4k full screen hanuman ji \#shorts \#shortvideo \#shortsvideo \#video \#viralvideo \#viralshorts \#viral \#status \#hanuman..." \newline
    2. "This is what happened to four year old Luz Gonzalez \#justiceforluz \#justiceforluzgonzalez \#justiciaparaluz..." \newline
    3. "Movietitle: Passenger 57 ...Security expert Carter (Wesley Snipes) is in a depressed state after a mishandling of a supermarket robbery leads to the death of his wife.. On the same flight, the famous terrorist Charles Lane (Bruce Payne) is escorted by the FBI to California to stand trial...." \\
    \midrule
    3 & 102 (15.6\%) & Inappropriate and offensive humor & \raggedright comedy, thank, use, funny, content, purpose, football, tip, gym, fair & 
    1. "You don’t mess with the ANATOLY...bodybuilder \#gym \#prank" \newline
    2. "between my two brothers i swear i actually have 3 dads lmfaoooo \#foryoupage \#fyp \#foryou \#viral \#comedy \#hilarious \#brother \#siblings \#siblingcheck \#siblinggoals" \newline
    3. "... white audiences are the least fun. THANK YOU CHAD ... \#mattrife \#improv \#crowdwork \#darkhumour" \\
    \midrule
    4 & 62 (9.5\%) & Pranks, games & \raggedright copyright, reel, chat, credit, warzone, forget, fun, intend, failarmy, disclaimer & 
    1. "try not to laugh \#fun \#funny \#prank \#snake \#scareprank \#fail \#failarmy \#foryou \#fyp \#trending \#funnyvideos \#failvideo \#learnontiktok" \newline
    2. "I Met Warzone’s MEANEST Kid \#shorts \#warzone2 \#warzone warzone warzone2..." \newline
    3. "You Sound Kind Of UGLY \#shorts \#warzone2 \#warzone warzone2..." \\
    \bottomrule
  \end{tabular}
  \caption{Latent Dirichlet allocation topic modeling for captions of problematic short videos (n = 651), showing topic summary, count, keywords, and example captions.}
\end{table*}

\subsection{Unsafe Short Videos: Danger Depicted}
RQ3 asked about the thematic topics in unsafe content. Before answering RQ3, we provide a thematic overview of six topics that emerged from the entire dataset. Short videos in the dataset encompass a wide range of topics, including daily vlogging (the number of captions with the largest topic distribution for this topic = 725/1676, 43.4\%), mental health- and physical health-related content (e.g., stigma awareness, travel; n = 353, 21.1\%), music and comedy (e.g., lyric captions; n = 150, 8.9\%), celebrities and influencers (e.g., dance, livestream; n = 113, 6.7\%), art showcase and self-advertisement (e.g., photography, handcraft; n = 220, 13.1\%), and how-do videos (e.g., cooking tips, gaming tutorials; n = 115, 6.9\%).

\begin{table*}[ht]
  \centering
  \begin{tabular}{p{0.04\textwidth}p{0.05\textwidth}p{0.07\textwidth}p{0.14\textwidth}p{0.57\textwidth}}
    \toprule
    \textbf{Topic \#} & \textbf{Count (\%)} & \textbf{Topic \newline Summary} & \textbf{Top Keywords} & \textbf{Example Captions} \\
    \midrule
    1 & 452 (54.3\%) & Fear-inducing content & \raggedright channel, content, best, hope, game, car, vs, license, usage, watch & 
    1. "Hatching a Children’s Pythons I will be at Animal Con USA this September..." \newline
    2. "My Everglades friends are being feisty tonight..." \newline
    3. "Hannah Landon- The Death of Bella Fontenelle \#greenscreen \#truecrime \#truecrimetiktok \#truecrimecommunity \#truecrimetok \#crime \#crimejunkie \#crimetok \#foryourpage \#foryoupage \#foryou \#fyp \#disturbing \#crimetiktok" \\
    \midrule
    2 & 144 (17.3\%) & Celebrity and influencers & \raggedright celebrity, provide, clip, update, daily, enjoy, ringtone, news, bts, music & 
    1. "Mary on a cross \#greysanatomy \#greysanatomyedits \#alexkarev \#meredithgrey \#mirandabailey \#georgeomalley" \newline
    2. "\#theuk07 rider \#influencer \#shotoniphone \#superbikes \#modified \#youtuber \#iphone \#14promax \#echo \#reelkarofeelkar..." \newline
    3. "Taylor Swift’s 4 MISHAPS At Eras Tour Tampa produced and edited by me using cap-cut editing software..." \\
    \midrule
    3 & 104 (12.5\%) & Mental health-related content & \raggedright train, cars, beamngdrive, drive, health, usa, uk, use, book, audiobook & 
    1. "You got this \#artisttoartist \#artistrelates \#artistoninstagram \#artistreels \#animeart \#animeartist \#artistlife \#animefan \#explore" \newline
    2. "...Take care of your mind and body. Love, Nawal \#anxietyhelp \#mentalhealth \#lessonlearned \#psychologyfacts \#humanbehavior \#awareness" \newline
    3. "\#600poundlife \#tlc \#obesity \#spaghetti \#2023 \#my600lblife \#newepisode I thought it was a serving bowl" \\
    \midrule
    4 & 133 (16.0\%) & Luxury lifestyle & \raggedright right, dog, latest, link, copyright, food, purpose, reel, click, thank & 
    1. "Let’s make a \$2k pizza for a celeb client..." \newline
    2. "My bridal morning Lingerie ... Katie Did \#vintageblogger \#vintageblog \#vintagevibes \#1950sfashion \#1950sstyle..." \newline
    3. "Boho small knotless braids with Human hair Stylist ... a link in my bio to book... \#nychairstylist \#floriahairstylist \#texashairstylist" \\
    \bottomrule
  \end{tabular}
  \caption{Latent Dirichlet allocation topic modeling for captions of short videos (n = 833), showing topic summary, count, keywords, and example captions.}
\end{table*}

\subsubsection{Problematic Content: Harm Framed as Entertainment}
Among 1,676 captions available for topic modeling, 651 captions were associated with problematic short videos. Problematic content features four topics (Table 3). The first topic features promotional content and advertisements for influencers and celebrities, often containing sexual content (n = 351/651, 53.9\%). Second, a large portion of the problematic content depicts unsettling entertainment content related to crime, true crime stories, emotional trauma, and depictions of physical harm (n = 136, 20.9\%). The third topic revolves around offensive humor, with hashtags such as \#darkhumor (n = 102, 15.6\%). While this content may be amusing to the adult audience, younger audiences may find it inappropriate. The last topic focuses more on pranks and gaming content, with hashtags such as \#scareprank and \#warzone (n = 62, 9.5\%). Overall, problematic videos blend nudity, violence, and harm with the frame of entertainment and commercialization. This trend may raise concerns about the implicit normalization of problematic content for children and teenagers.

\subsubsection{Mentally Distressing Content: Anxiety-triggering Narratives}
Among 1,676 captions available for topic modeling, 833 captions were associated with mentally distressing short videos. Four topics are identified (Table 4). The first topic depicts a wide range of fear-inducing content, such as frightening animals and movie clips with unsettling plots (n = 452/833, 54.3\%). Such fear-evoking content can be distressing to children and teenagers in terms of both visuals and content. The second topic centers on celebrities and influencers (n = 144, 17.3\%). In particular, makeup influencers can deliberately or unintentionally promote unrealistic beauty standards that encourage social comparison, potentially leading to appearance-related anxiety and decreased self-esteem. The third topic is related to mental health, with life challenges, therapy, and emotional struggles (n = 104, 12.5\%). Even though this content often aims to raise awareness, it may unintentionally trigger anxiety and depressive thoughts among children and teenagers by depicting symptoms and struggling experiences that resonate deeply with the audience. For example, short videos promoting mental health awareness may inadvertently evoke distress by reinforcing negative emotional states. The last topic covers vlogging of luxury lifestyles (n = 133, 16.0\%), where influencers showcase extravagant living, vintage aesthetics, or glamorous parties. These narratives can create unrealistic life expectations for children and teenagers who may not have a full understanding of life realities and complexities. This disconnect may induce anxiety or a sense of inadequacy. Overall, mentally distressing content mostly does not intend to cause mental distress, but this content may implicitly trigger anxiety and negative emotions among children and teenagers with cognitive and emotional capacities still in development.

\section{Discussion}
Using multimodal feature analysis and topic modeling of short videos recommended to children and teenagers on Instagram Reels, TikTok, and YouTube Shorts, this study identified audiovisual and thematic markers of unsafe (i.e., problematic and mentally distressing) short videos. We highlight two major findings in this study. First, the distinct features of unsafe short videos manifest through both harmful video content and multimodal features, such as darker visuals. Second, by expanding the analysis of unsafe content to both problematic and mentally distressing content, we identified three distinctive ways unsafe content can produce harm: (a) explicit harm through inappropriate content, (b) implicit harm framed as entertainment and commercialization, and (c) unintentional harm embedded in mental health-related content.

Our findings suggest that darker visuals of unsafe content add another dimension to its harmful effect on children and teenagers. Although unsafe short videos are manually annotated based on the video content, the dark and gloomy visual features suggest an additional layer of harm. Research in color and psychology suggests that brightness and saturation are often positively associated with individual emotional arousal and engagement, with darker tones linked to negative emotional states and decreased motivation \cite{berger_arousal_2011, wilms_color_2018}. In particular, gray hues are often associated with anxiety and low energy \cite{carruthers_manchester_2010}. Further, the activating role of bright colors explains the differentiated association between content types and video engagement. Child-directed short videos with bright and saturated visuals are more likely to activate emotional arousal and elicit higher levels of engagement on short video platforms. On the contrary, unsafe content with darker visuals receives lower levels of engagement, in terms of the number of views, likes, and comments. In particular, unsafe content directed to children possessed a faster tempo and gloomier visual features. However, unsafe content may still attract other forms of engagement, such as dislikes, though this data is not available in the current dataset. Therefore, beyond explicit harmful video content, anxiety-inducing visual elements in these videos may subtly influence the well-being of children and teenagers and are associated with less engagement.

\subsubsection{A Typology of Online Harm} Our findings on thematic markers of unsafe content introduce a topology of online harm to children: (a) \textit{explicit harm} through inappropriate content, (b) \textit{implicit harm} framed as entertainment and commercialization, and (c) \textit{unintended harm} embedded in mental health-related content. The harm of unsafe content may arise not only from explicit or violent content but also from harmful narratives framed as entertainment or commercials. We observe a wide range of short videos depicting crime-related movie clips and comedies, which may introduce a subtler form of risks to children and teenagers who are learning and imitating the social environment, according to social cognition theory \cite{bandura_social_2001}. Although crime stories and dark humor are entertaining, they may normalize harmful themes and damage the well-being of children and teenagers.

Additionally, ordinary content such as mental health awareness or luxury vlogging can have unintended impacts. Mental health-related content often aims to promote awareness and well-being, but its depiction of struggling experiences and negative emotions may inadvertently resonate and trigger the audience’s traumatic experiences and anxiety  \cite{hancock_im_2008}. Similarly, vlogs that portray an idealized, luxurious lifestyle may create unrealistic expectations and a sense of inadequacy among children and teenagers \cite{chae_explaining_2018, tian_antecedents_2023}. This type of content often blends with implicit brand integration and product promotion, which aligns with the previous finding that product placements on child-directed YouTube videos are often implicit without proper disclosure \cite{choi_brand_2023}. This raises important ethical concerns because children are particularly vulnerable to advertising effects \cite{auty_exploring_2004}. These findings highlight the implicit dangers of the seemingly harmless content on short video platforms for children and teenagers who may struggle to critically evaluate the content.

The findings do not mean that such content should be removed altogether since it is appropriate and even entertaining to the majority of adult audiences. The key lies in the ongoing development of cognitive and emotional capabilities among children and teenagers, which makes them particularly vulnerable to such content. While social media and short video platforms do restrict children under 13 due to the Children’s Online Privacy Protection Act (COPPA), which prohibits websites from collecting information on children younger than 13 years without parental permission, the unfortunate reality is that a functional age verification system is still lacking. As a result, many children and teenagers have access to social media platforms and are exposed to social media content that can be problematic or mentally distressing to children and teenagers in their developmental stages. Overall, this study introduces a useful typology for future research on online harms, providing a framework to better understand and address the evolving risks posed by short video content.

\subsection{Practical Implications}
Our study highlights the pressing need to reduce children’s and teenagers’ exposure to unsafe content on social media, providing actionable implications for algorithmic content moderation and regulatory interventions.

First, the study’s findings call for scaling up the detection and mitigation of unsafe content with algorithmic content detection and moderation systems based on unique audiovisual and thematic features to address explicit, implicit, and unintended harms. Beyond detecting overtly harmful content, audiovisual markers, such as darker visuals and gloomy tones, should be integrated into content moderation algorithms to flag potentially harmful content more effectively. By incorporating these markers, platforms can move beyond text-based detection systems and proactively moderate content that may pose subtle psychological risks to children and teenagers. Further, our findings call for adaptive algorithm systems that prioritize safe and appropriate content that aligns with their developmental needs and contributes to a safer digital environment. However, this suggestion does not mean audiovisual features should be the sole criterion for online content moderation. Rather, our findings illustrate that audiovisual features can act as useful indicators to assist with unsafe content removal. Effective content moderation should strike a careful balance between promoting online safety and protecting freedom of expression, ensuring a safe and inclusive digital environment for all users.

Second, our findings call for regulatory and legislative efforts to establish clear guidelines on age verification on social media platforms, ensuring that children and teenagers under 13 have no access to unsafe content. Given the broad spectrum of online harm, robust age verification mechanisms are essential to prevent children and teenagers from bypassing restrictions and getting exposed to unsafe content during critical developmental stages.

\subsection{Limitations and Future Directions}
This study is not without limitations. First, the multimodal feature analysis relies on 16 basic multimodal features, which may not fully capture the audiovisual elements in short videos. Future research may expand the scope of this study and incorporate more platform-specific features and filters to provide a more comprehensive understanding of multimodal features of unsafe content.

Second, the generalizability of the study’s findings is limited by the dataset in two ways: (a) some short videos became unavailable during the study due to takedowns resultant from internal content moderation or changes in privacy settings, and (b) thematic analysis was conducted only on English-language captions. On the first point, only 4.1\% (n = 186/4,496 URLs) of videos were inaccessible. Because the purpose of the study was to identify multimodal features of unsafe videos that may lead to subsequent negative impacts over time, we think this limitation is a minor point. On the second point, only 5.5\% (n = 93/1,696 captions) of the captions are in non-English languages. Future work needs to take into consideration all languages in texts, and culture- or language-specific multimodal features to expand the global relevance and generalizability of this work \cite{lai2021cultural}.

Third, although this study analyzes short video content within the context of an algorithm auditing experiment, this study is observational in nature. As such, it provides correlational evidence linking multimodal features of short videos to unsafe content, without addressing the question of how such features influence subsequent information processing and psychological or behavioral changes. Future research can build on these findings using longitudinal designs or experimental approaches to better understand how unsafe multimodal contents evolve and affect children and teenagers.

Lastly, engagement with short videos was measured as the number of views, likes, and comments, but the data on dislikes or downvotes was unavailable. This is because only YouTube Shorts allows dislikes but the platform does not report them separately. The like count reflects the net value after subtracting dislikes. This lack of negative engagement metrics limits our ability to capture more nuanced audience responses, especially for unsafe content that is widely viewed but potentially negatively received. Future research could incorporate alternative metrics, such as watch durations and shares, to better evaluate audience engagement at the individual level.

\section{Conclusion}
Relying on the feature-level and content-level analysis of 4,492 short videos recommended to children, this study reveals that unsafe short videos possess darker visuals and include both harmful content framed as entertainment and harm-triggering ordinary content. We introduce a useful framework of three forms of online harm to children (i.e., explicit, implicit, unintended), providing a unique lens for understanding the dynamic, multifaceted online risks facing children and teenagers in critical developmental stages. Our findings have actionable implications for algorithmic content moderation and regulatory interventions, contributing to a safer digital environment for children and teenagers.

\section{Declaration of Competing Interest}
The authors declare the following financial interests/personal relationships which may be considered as potential competing interests: Jingwen Zhang and Drew Cingel report a relationship with ongoing social media litigation efforts that include paid consulting or advisory. All other authors declare that they have no known competing financial interests or personal relationships that could have appeared to influence the work reported in this paper.


\vspace{120pt}
\appendix
\section{Appendix}

The association between multimodal features and short video content category across platforms.

\begin{figure}[h]
\centering
\includegraphics[width=0.95\columnwidth]{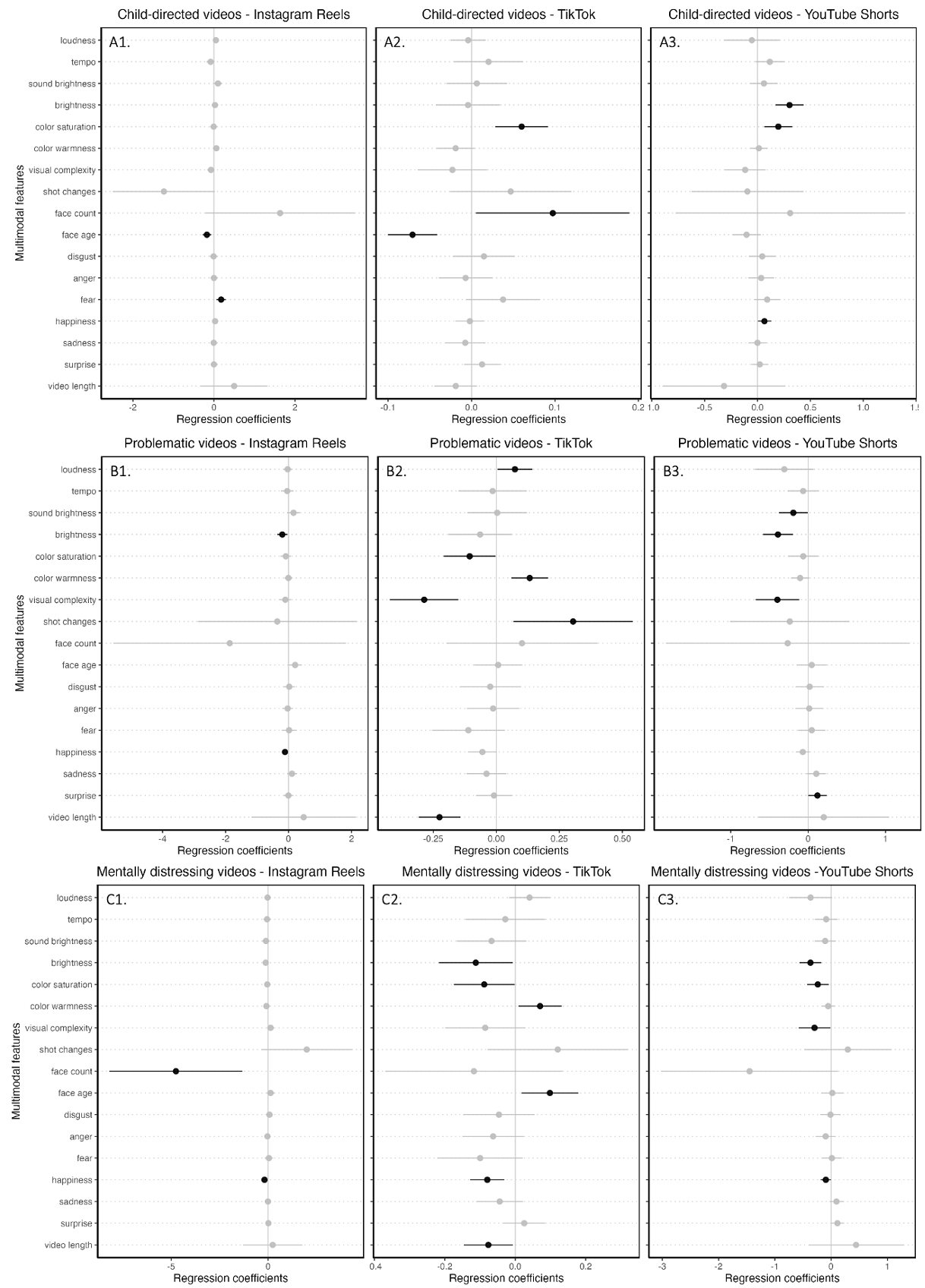}
\caption{Regression coefficients and confidence intervals of multimodal features on the extent to which short videos on (1) Instagram Reels, (2) TikTok, and (3) YouTube Shorts are (A) child-directed, (B) problematic, or (C) mentally distressing. Multimodal features are normalized for comparison.}
\label{figure_appendix}
\end{figure}

\end{document}